\documentclass[a4paper]{jpconf}
\usepackage{graphicx}
\pdfoutput=1
\usepackage{comment}
\usepackage[T1]{fontenc}

\usepackage{units}
\usepackage{booktabs}
\usepackage{tabularx}
\usepackage{hyperref}
\usepackage{xcolor}

\newcommand{\PT}           {\ensuremath{p_{\rm T}}}

\newcommand{\ppi}          {\ensuremath{{\rm p}/\pi}}

\newcommand{\lks}          {\ensuremath{{\rm \Lambda}/\rm{K}^{0}_{S}}}

\begin{document}

\title{Transverse Momentum Distributions of Identified Particles in p--Pb Collisions at \boldmath$\mathrm{\sqrt{s_{NN}} = 5.02\ TeV}$}

\author{J. Anielski}

\address{Institut f\"ur Kernphysik, Universit\"at M\"unster, Wilhelm-Klemm-Str. 9, 48149 M\"unster, Germany}

\ead{j.anielski@wwu.de}

\begin{abstract}

Transverse momentum ($p_{\mathrm{T}}$) distributions of identified hadrons produced in p--Pb collisions at $\mathrm{\sqrt{s_{NN}} = 5.02\ TeV}$ have been measured at mid-rapidity (0  $< y_{\mathrm{CMS}} <$  0.5) by ALICE. Particle tracks are reconstructed using the central barrel detectors. Particle identification is performed via specific energy loss, time-of-flight or their characteristic decay topology over a wide transverse momentum range (0 GeV/\textit{c} up to 8 GeV/\textit{c}).

Spectral shapes and particle ratios are measured in six multiplicity classes. They are compared with several model calculations and results from Pb--Pb collisions at $\mathrm{\sqrt{s_{NN}} = 2.76\ TeV}$ and pp collisions at $\mathrm{\sqrt{s_{NN}} = 7\ TeV}$ at the LHC. The results are discussed with respect to possible collective effects in p--Pb collisions.

\end{abstract}

\section{Introduction}

Recent measurements of di-hadron correlations in p--Pb collisions at $\mathrm{\sqrt{s_{NN}} = 5.02\ TeV}$ revealed a double-ridge pattern, reminiscent of the one observed in Pb--Pb collisions~\cite{Abelev:1497210,CMS:2012qk,Aad:2012gla,Khachatryan:2010gv}. This raises the question of the possible existence of collective effects in high multiplicity p--Pb collisions. Further insight into the observed phenomena can be gained by studying the evolution of spectral shapes with the particle mass and particle ratios as a function of charged-particle density. 

Primary charged particles ($\pi\mathrm{^{\pm}, K^{\pm}, p\ and\ \bar{p}}$) are identified by their specific energy loss ($\mathrm{d}E/\mathrm{d}x$) in the Inner Tracking System (ITS) and the Time Projection Chamber (TPC)~\cite{Alme:2010ke} and by their time-of-flight with the TOF detector~\cite{Akindinov:2013tea}. Weakly decaying particles ($\mathrm{ K^{0}_{s}, \Lambda\ and\ \bar{\Lambda}}$) are identified by their characteristic decay topology.  In this work, we present results for primary particles, defined as all particles from the collision, excluding products of weak decays of strange particles. 
The transverse momentum distributions have been reported in~\cite{ALICE:2013pPbSpectra}. The particle identification techniques are explained in detail in~\cite{prl-spectra, Abelev:2013vea, ALICE:2013xaa}.


\section{Particle ratios as a function of \boldmath$p_{\mathrm{T}}$ and d$N_{\mathrm{ch}}/\mathrm{d}\eta$}

\begin{figure}
\begin{center}
\includegraphics[trim=0cm 0cm 0cm 1.2cm, clip=true, width=0.49\textwidth] {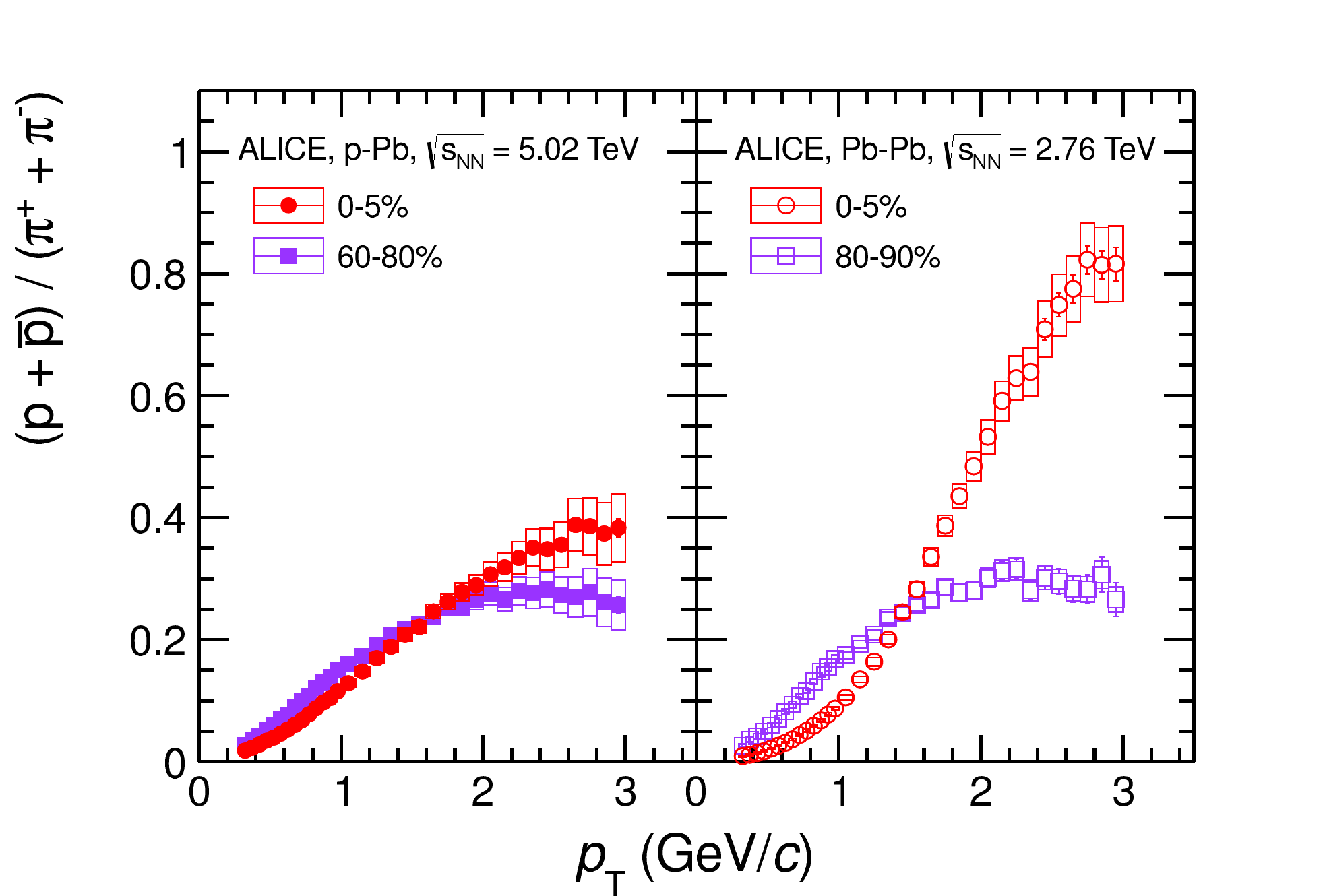}
\includegraphics[trim=0cm 0cm 0cm 1.2cm, clip=true, width=0.49\textwidth] {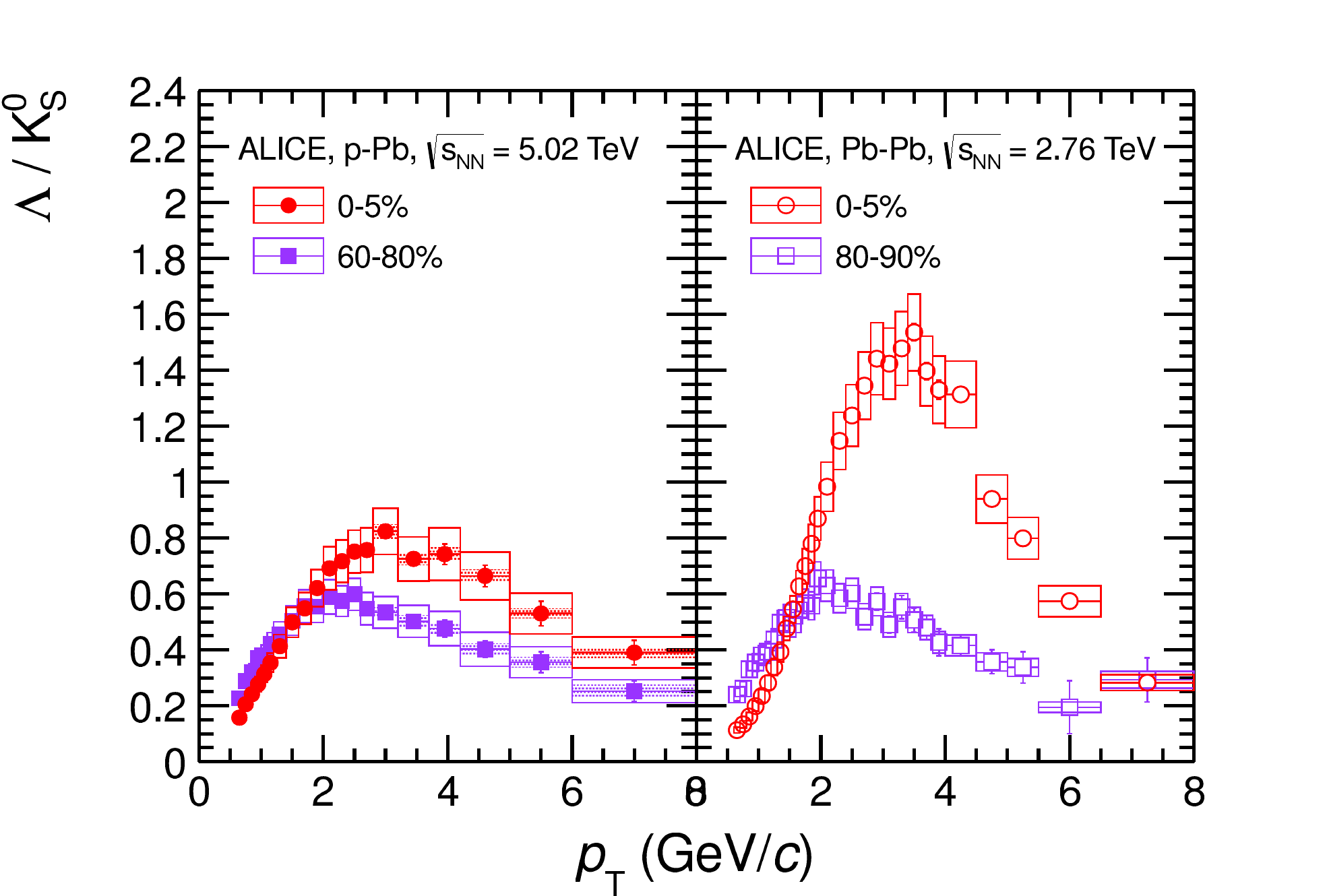}
\end{center}
\caption{\label{fig:ProtonPiRatio} p/$\mathrm{\pi}$ and $\mathrm{\Lambda/K^0_s}$ ratio as a function of \PT{} for p--Pb and Pb--Pb collisions. Systematic errors are largely correlated across multiplicity. Multiplicity uncorrelated errors are drawn as a band for p--Pb. From~\cite{ALICE:2013pPbSpectra}.}
\end{figure}

The ratios p/$\mathrm{\pi}$ and $\mathrm{\Lambda/K^0_s}$ as a function of transverse momentum are shown in Fig.~\ref{fig:ProtonPiRatio} for a high and a low multiplicity class in p--Pb and in Pb--Pb collisions. For high multiplicity events, an increase of protons at intermediate \PT{} and a corresponding depletion at low \PT{} are evident. The change of shape with multiplicity can be observed for protons (and $\mathrm{\Lambda}$) in both collisions systems, but the relative increase in multiplicity and the absolute magnitude of the effect, however, are much smaller in p--Pb collisions. In Pb--Pb collisions this effect is generally attributed to collective flow or quark recombination~\cite{Fries:2003vb, Bozek:2011gq,Muller:2012zq}. 

\begin{figure}
\begin{center}
\includegraphics[trim=0cm 0cm 0cm 1.2cm, clip=true, width=0.49\textwidth] {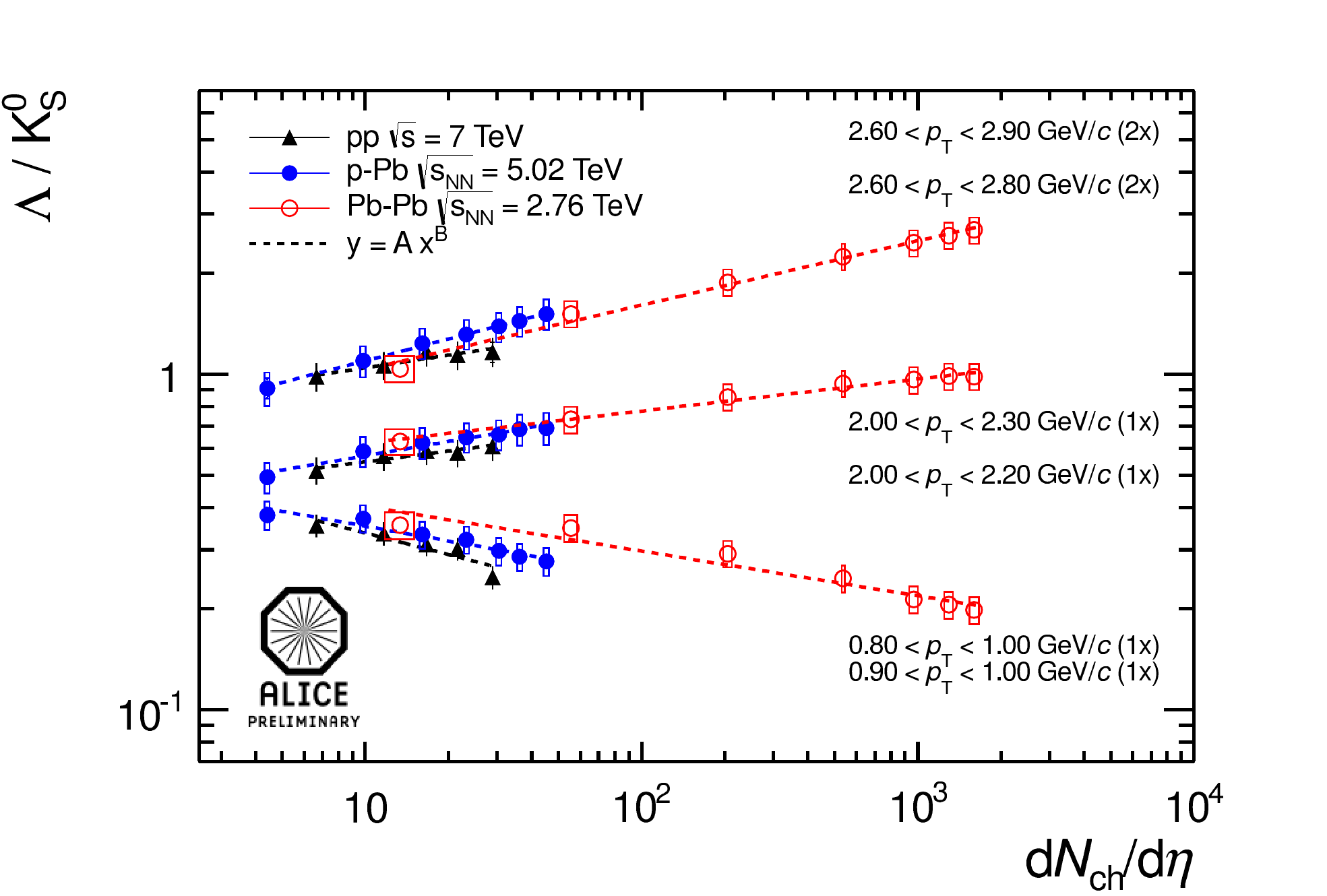}
\includegraphics[trim=0cm 0cm 0cm 1.2cm, clip=true, width=0.49\textwidth] {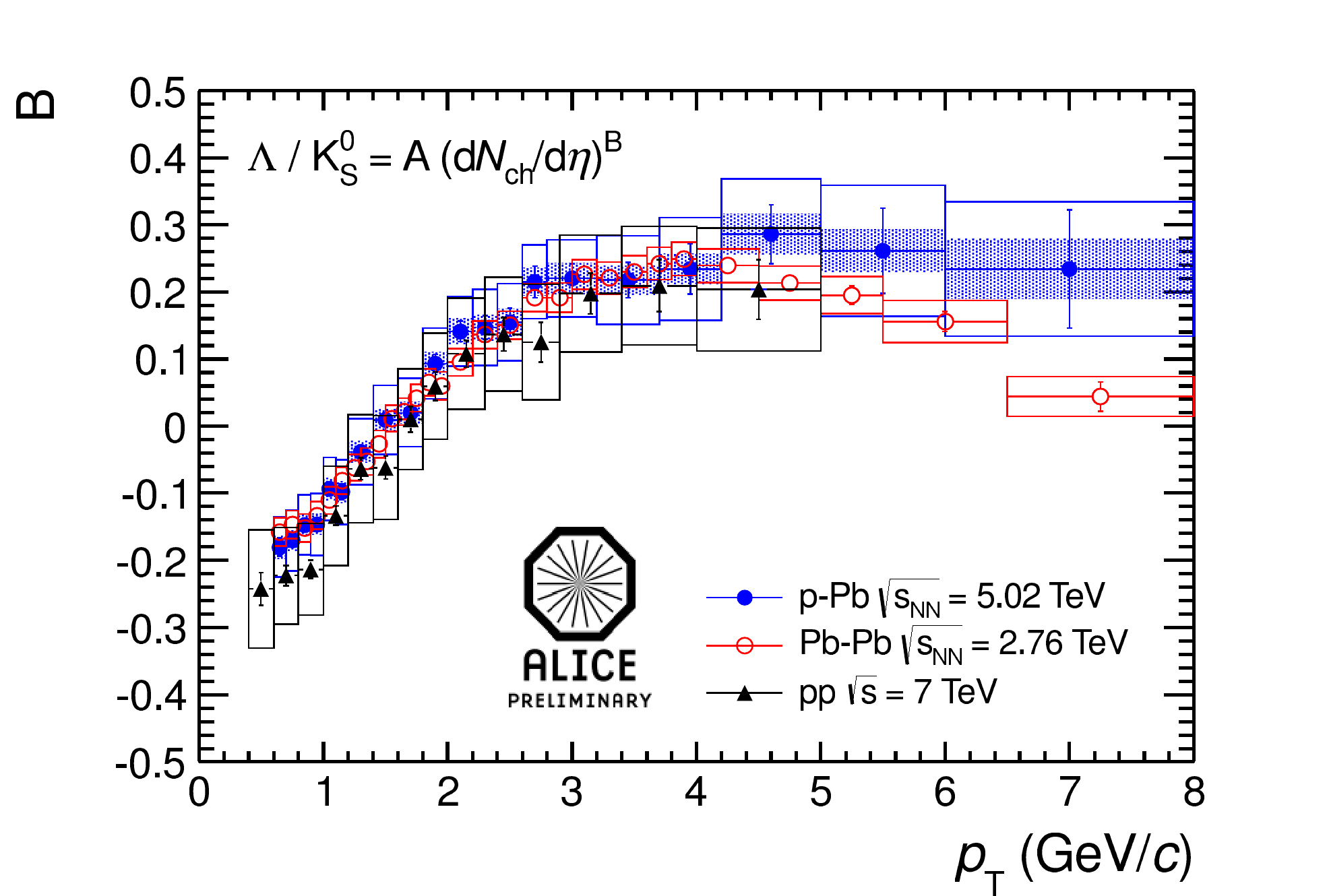}
\end{center}
\caption{\label{fig:scaling}Fit of the $\mathrm{\Lambda/K^0_s}$ ratio as a function charged multiplicity with power law for three different \PT{} slices and for pp, p-Pb and Pb-Pb (left). Extracted exponents B as a function of \PT{} (right). Based on~\cite{ALICE:2013pPbSpectra}.}
\end{figure}

To further quantify this observation, we plot the $\mathrm{\Lambda/K^0_s}$ ratio as a function of charged-particle multiplicity for three different momentum bins and for pp, p--Pb and Pb--Pb collisions in Fig.~\ref{fig:scaling} on the left. For each momentum slice, the particle ratios follow a similar trend for the three collision systems and are fitted individually with a power-law. The exponents of the fits, plotted as a function of \PT{} in Fig.~\ref{fig:scaling} on the right, show similar values in all collision systems and follow the same trend. This observation indicates a universal behavior of the \lks{} ratio as a function of multiplicity, regardless of the collision system. A similar scaling behavior applies to the p/$\mathrm{\pi}$ ratios.

\section{Global Blast-Wave Fit}

\begin{figure}
  \begin{center}
    \includegraphics[trim=0cm 0cm 0cm 1.2cm, clip=true, width=0.5\textwidth]{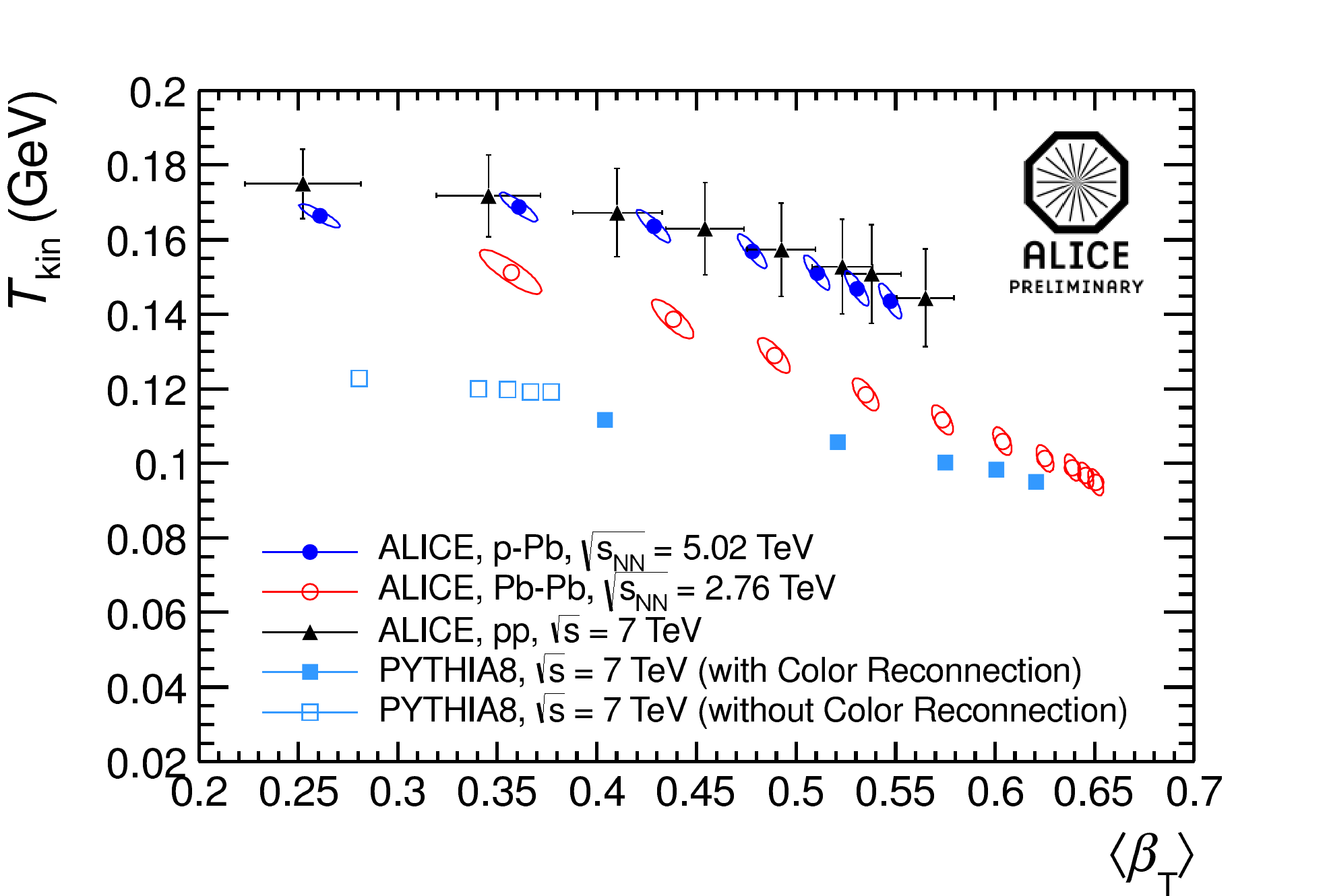}
  \end{center}
  \caption{\label{fig:blastwave}Fit parameters from combined Blast-Wave fit for several multiplicity/centrality classes in pp, p--Pb and Pb--Pb. Charged-particle multiplicity increases from left to right. Based on~\cite{ALICE:2013pPbSpectra}.}
\end{figure}

To compare the evolution of the spectral shape with multiplicity between different collision systems a simultaneous blast-wave fit to the $\pi$, K, p and $\Lambda$ spectra was performed for each multiplicity bin. The blast-wave framework assumes a locally thermalized and collectively expanding medium and a common freeze-out for all particles~\cite{Schnedermann:1993ws}. It is known, as discussed e.g. in~\cite{prl-spectra}, that the fit parameters depend significantly on the fitting range. Nevertheless, the blast-wave framework is a useful tool to describe spectra with a small set of parameters. The extracted freeze-out temperatures ($T\mathrm{_{kin}}$) and the transverse flow ($\beta_{T}$) show a similar behavior in p--Pb and Pb--Pb collisions (Fig.~\ref{fig:blastwave}). This trend is generally understood in Pb-Pb collisions as due to an increasingly stronger radial flow as a function of centrality  in heavy-ion collisions. To further test this  picture, a simultaneous blast-wave fit to transverse momentum distributions in pp collisions as a function of multiplicity has been performed. Due to the limited available range, the fit has been performed for $\pi$, K and p for 0.5-1.0, 0.3-1.5 and 0.5-2.5 GeV/\textit{c}, respectively. These are slightly different from the ranges used for p--Pb~\cite{ALICE:2013pPbSpectra}. It should also be noted that the pp data have been divided in multiplicity intervals based on charged tracks at mid-rapidity, which could introduce a bias on the shape of the spectra. The extracted parameters show a very similar trend to the p--Pb points, within the caveats mentioned above.

Despite this suggestive observation, it has to be noted that effects other than radial flow can produce the observed trend. To illustrate this, pp spectra generated with PYTHIA~8 (tune~4C) have been fitted for different multiplicity slices. The open symbols show the results for PYTHIA without the color reconnection (CR) mechanism~\cite{Skands:2007zg,Schulz:2011qy}. For the spectra with CR (closed symbols) a trend qualitatively similar to the one observed in pp, p--Pb and Pb--Pb collisions can be seen.

\section{Comparison with Models}

\begin{figure}

  \begin{center}
  \includegraphics[trim=0cm 0cm 0cm 1.2cm, clip=true, width=0.5\textwidth]{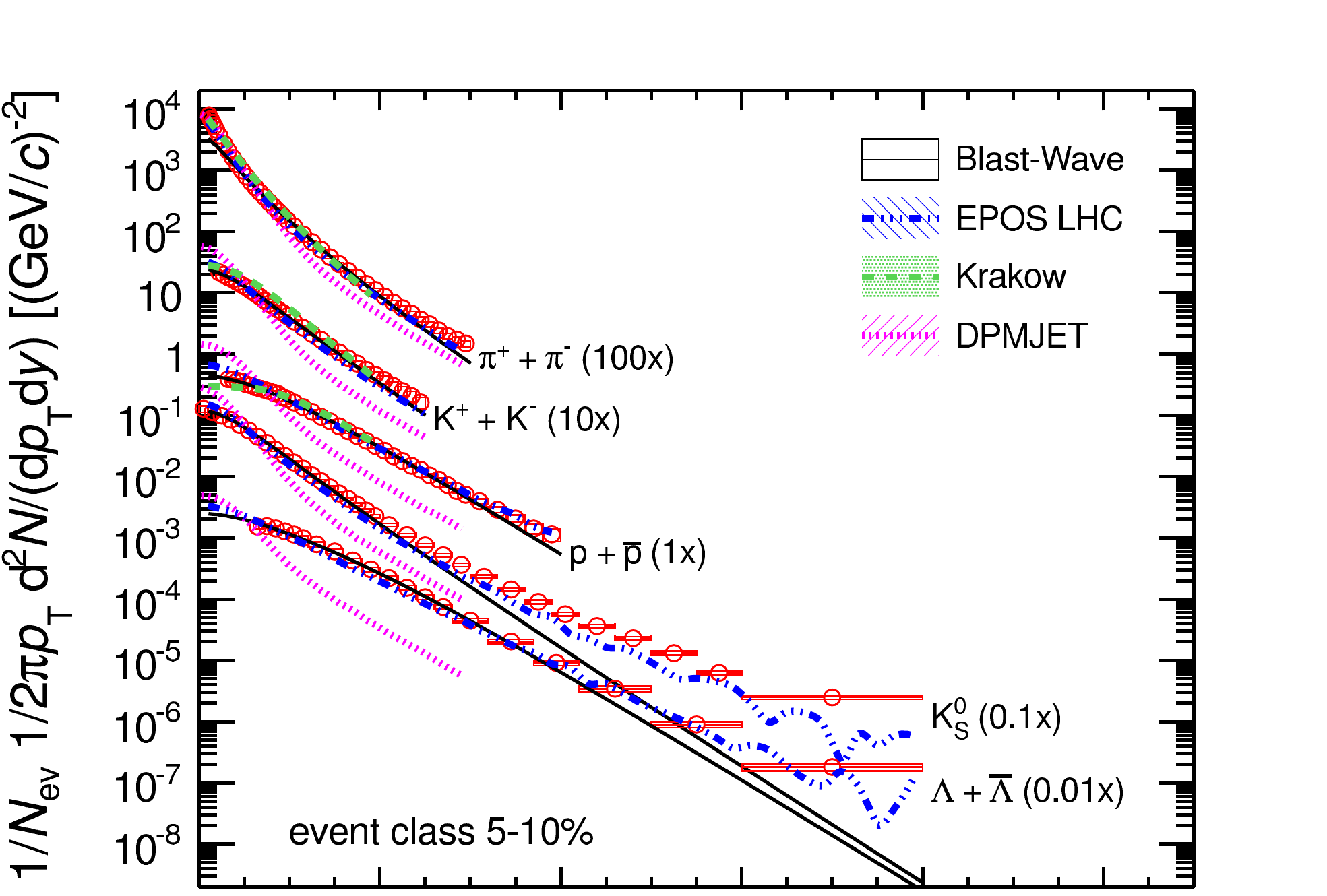}  

  \vspace{-0.08cm}

  \includegraphics[trim=0cm 0cm 0cm 1.2cm, clip=true, width=0.5\textwidth]{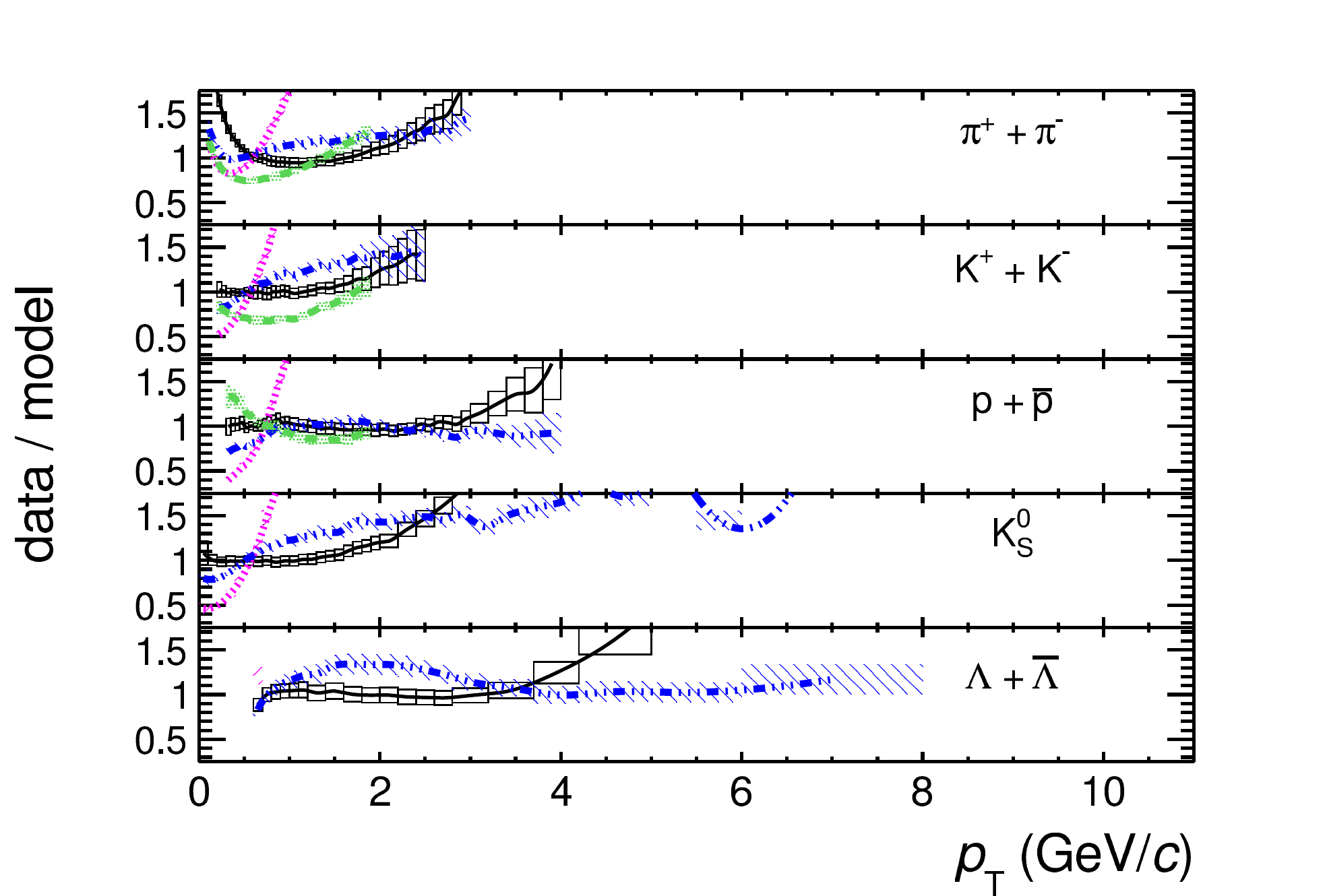}  

  \end{center}
  
  \caption{\label{fig:models} Comparisons of the spectra in the 5-10\% multiplicity bin with several models. From~\cite{ALICE:2013pPbSpectra}.}

\end{figure}

The measured \PT{} distributions in the 5-10\% bin are compared with calculations from DPMJET~\cite{Roesler:2000he}, Krak\'{o}w~\cite{Bozek:2011if}, and EPOS LHC 1.99 v3400~\cite{Pierog:2013ria} in Fig~\ref{fig:models}. DPMJET is a QCD-inspired generator, which treats soft and hard scattering processes in an unified way. It can reproduce the pseudorapidity distribution of p--Pb collisions at the LHC~\cite{ALICE:2012xs}, but it fails to describe the transverse momentum spectra of identified particles. The hydrodynamical Krak\'{o}w model uses a Glauber model based on Monte Carlo to create fluctuating initial conditions. It can reproduce the shape of the measured pion and kaon spectra reasonably well up to 1~GeV/\textit{c}, but fails for higher momenta. This could, in the hydrodynamical framework, indicate the onset of a non-thermal component. The EPOS  model is founded on parton-based Gribov Regge theory and implements an hydrodynamical evolution for bulk particles. It can describe the pion and proton distributions within 20\% over the full measured range, but shows larger deviations for kaons and lambdas.

\section{Discussion}

The measurement of identified particle spectra adds valuable information to address the presence of possible collective effects in p--Pb collisions. The \ppi{} and \lks{} ratios show a similar behavior in p--Pb and Pb--Pb collisions at the LHC and the increase of the ratio scales with the charged-particle multiplicity for pp, p--Pb and Pb-Pb collisions. The blast-wave analysis and the comparison with models indicate that final state effects seem to be needed to reproduce the measured spectra. The blast-wave analysis of Monte Carlo spectra, generated with PYTHIA, shows that other effects like color-reconnection can produce a pattern similar to radial flow.

\section*{References}

\bibliographystyle{apsrev4-1}
\bibliography{SQMprod_jonas}

\end{document}